# Ultrahigh toughness polycrystalline ceramics without fading of strength

Koji Matsui[1,2*], Kouhei Hosoi[1,3], Bin Feng[1,2], Hidehiro Yoshida[1,4] and Yuichi Ikuhara[1,2,5]

[1*]Next Generation Zirconia Social Cooperation Program, Institute of Engineering Innovation, The University of Tokyo, Bunkyo-ku, Tokyo 113-8656, Japan

[2]Institute of Engineering Innovation, The University of Tokyo, Bunkyo-ku, Tokyo 113-8656, Japan

[3]Inorganic Materials Research Laboratory, Tosoh Corporation,4560, Kaisei-cho, Shunan, Yamaguchi 746-8501, Japan

[4]Department of Materials Engineering, School of Engineering, The University of Tokyo, 7-3-1 Hongo, Bunkyo-ku, Tokyo, 113-8656, Japan

[5]Nanostructures Research Laboratory, Japan Fine Ceramics Center, 2-4-1 Mutsuno, Atsuta, Nagoya 456-8587, Japan

Correspondence and requests for materials should be addressed to K. M. (email: kojimatsui@g.ecc.u-tokyo.ac.jp)

**Zirconia ceramics have been known as a structural material with high fracture toughness ($K_{Ic}$) and strength ($\sigma_f$) since Garvie *et al*. discovered phase transformation toughening in 1975 [1]. Although these mechanical properties are the most excellent among the advanced ceramics, the $K_{Ic}$ has not yet reached the level of metallic materials. Here, we demonstrate** **for the first time that 2.9 mass% $Y_2O_3$-stabilized $ZrO_2$s doped with $Al_2O_3$ greatly exceed the $K_{Ic}$ of conventional zirconia ceramics which is comparable to those of metallic materials, even with slightly higher $\sigma_f$. The excellent mechanical properties in the proposed ceramic materials will be useful for further expanding the application of advanced ceramics to many engineering fields.**

Among the advanced ceramics, polycrystalline tetragonal $ZrO_2$ with excellent mechanical properties has been the most extensively commercialized for the use of optical fiber connectors, grinding media, precision parts, and dental applications [2-4]. The high $K_{Ic}$ of the tetragonal $ZrO_2$ is attributed to the transformation toughening mechanism, where crack propagation resistance is increased by the stress-induced tetragonal-to-monoclinic (T→M) phase transformation [5] accompanied with an expansion of the unit cell volume of approximately 4 vol%. Although the mechanical properties above are the most excellent among the advanced ceramics, the $K_{Ic}$ has not yet reached the level of metallic materials [6]. To enhance the $K_{Ic}$, it is effective if the T→M phase transformation could be induced with lower stress. When doped with low concentration of stabilizer into $ZrO_2$, the T→M phase transformation easily occurs with a low stress because stability of the tetragonal phase is reduced. However, it has also been difficult to obtain the fully densified sintered body without cracks so far, because the T→M phase transformation spontaneously occurs even without any stress in the cooling process of sintering. For example, in $Y_2O_3$-stabilized $ZrO_2$, it has been reported that using the raw powder synthesized by a thermal decomposition process, the $K_{Ic}$ of sintered body obtained by pressureless sintering at 1500°C was drastically reduced at $Y_2O_3$ concentrations < 3.6 mass% [7]. This result suggests that because the grain size of sintered body is large when sintering at the high temperature with low $Y_2O_3$-concentration, the stability of the tetragonal phase is reduced, and cracks are consequently generated by the spontaneous T→M phase transformation which causes a decrease in $K_{Ic}$. Since the sinterability of the raw powder strongly depends on uniformity of the particle structure and chemical composition, it is essential to design the raw powders with low stabilizer concentration, which can be sintered at a low temperature without forming cracks.

Among varies powder-synthesis methods, the hydrolysis process has an advantage that raw powders can be stably produced with high quality by exact control of the particle structure [8]. Furthermore, the stabilizer is uniformly dissolved in the particle interior [9]. Recently, we have successfully developed a new $ZrO_2$ powder with low concentration of stabilizer, i.e., 2.9 mass% $Y_2O_3$-stabilized $ZrO_2$ doped with 0.25 mass% $Al_2O_3$ (ZA) powder using a refined hydrolysis process [8], which can be fully densified by pressureless sintering at 1250°C in air [10]. The powder doped with 0.25 mass% $Al_2O_3$ greatly enhances the densification during sintering, as will be described later. In this work, we investigated the sintering behavior, microstructure, $K_{Ic}$ and $\sigma_f$ of ZA sintered bodies. It is known that when the hot isostatic pressing (HIP) treatment was conducted to conventional $Y_2O_3$-stabilized $ZrO_2/Al_2O_3$ composites with the $Al_2O_3$ concentrations ≤ 20 mass%, the $\sigma_f$ increases with increasing $Al_2O_3$ concentration [11]. Therefore, two kinds of powders, namely 2.9 mass% $Y_2O_3$-stabilzed $ZrO_2$s containing 1 mass% $Al_2O_3$ (Z1A) and 20 mass% $Al_2O_3$ (Z20A) were prepared in order to confirm the HIP effect. The $K_{Ic}$ and $\sigma_f$ were examined at ZA, Z1A, and Z20A bodies HIPed at 1350, 1350, and 1500°C under a pressure of 150 MPa in Ar, respectively.

The starting material used was ZA. The raw powder was press-formed, and the green bodies were subsequently sintered at temperatures of 1000-1500°C for 2 h in air. Figure 1(a) shows the relative density ($\rho$) and average grain size of the sintered bodies as a function of the sintering temperature. The $\rho$ of ZA reached 99.8% at 1250°C, rapidly decreased above 1375°C, and became an almost stable value of 95% at 1400-1500°C. The average grain size gradually increased up to 1400°C, and then rapidly increased when the temperature reached 1500°C. Scanning electron microscopy (SEM) images of ZA sintered at 1350 and 1400°C are shown in Fig. 1(b). In the ZA body sintered at 1350°C, the average grain size was 0.32 μm. At 1400°C, the average grain size increased to 0.40 μm and cracks were observed. The factor of crack generation is the volume expansion caused by the T→M phase transformation, and the $\rho$ consequently decreased to approximately 95 %. According to the phase diagram for the $ZrO_2$-$Y_2O_3$ system [12], the crystal phase is the single tetragonal phase in the range of sintering temperature in the present experimental for ZA. Therefore, the decrease in $\rho$ that appeared above 1375°C can be understood by the formation of monoclinic phase generated at the cooling stage of sintering. The microstructure in ZA body sintered at 1350°C was analyzed with scanning transmission electron microscopy (STEM)- energy dispersive X-ray spectroscopy (EDS) technique. Fig. 1(c) shows a typical STEM image and the corresponding element mapping images for Zr-$K\alpha$, Y-$K\alpha$, and Al-$K\alpha$. As evident from the Y-$K\alpha$ mapping image, $Y^{3+}$ ions were detected over all the grains and the distribution was almost homogeneous. $Al^{3+}$ ions, in contrast, were clearly segregated along the grain boundaries. The densification of $Y_2O_3$-stabilized $ZrO_2$ has been reported to be enhanced by doping of a small amount of $Al_2O_3$ [13, 14]. In this way, we confirmed that the ZA body sintered at 1350°C consists of tetragonal grains with a homogeneous $Y^{3+}$ ion distribution and $Al^{3+}$ ions segregate along grain-boundaries.

Mechanical properties of ZA were examined by three-point bending test for $\sigma_f$ and indentation fracture (IF) method for $K_{Ic}$. As shown in Fig. 2, the $\sigma_f$ and $K_{Ic}$ of ZA sintered at 1350°C were 1190 MPa and 23 MPa·m$^{0.5}$, respectively. Furthermore, mechanical properties of ZA, Z1A, and Z20A bodies treated with HIP were also examined at the same methods as above (Fig. 2). The $\sigma_f$ and $K_{Ic}$ of ZA body HIPed at 1350°C were 1340 MPa and 22 MPa·m$^{0.5}$, respectively. By the 1350°C-HIP treatment, the $\sigma_f$ was higher than that of ZA obtained by pressureless sintering at the same temperature. According to the report in Tsukuma *et al*. [15], because the HIP treatment is carried out under a high pressure in Ar, densification progressed easier compared to pressureless sintering, and consequently the size of flaw that causes a decrease in strength becomes smaller. Therefore, the increase in $\sigma_f$ can be understood by a decrease in flaw size. In Z1A body HIPed at 1350°C, $\sigma_f$ is 1510 MPa and $K_{Ic}$ is 22 MPa·m$^{0.5}$. Comparing Z1A and ZA HIPed, it can be seen that the $\sigma_f$ increased by the increase in $Al_2O_3$ concentration. On the other hand, the $K_{Ic}$ changed little for ZAs sintered in air, ZA and Z1A HIPed samples. In the Z20A containing excess $Al_2O_3$, at the 1500°C-HIP treatment, $\sigma_f$ =1980 MPa and $K_{Ic}$=16 MPa·m$^{0.5}$. The $\sigma_f$ of Z20A greatly enhanced, but the $K_{Ic}$ was lower than those of ZA and YZ1A. This can be understood as follows.

Dispersion of $Al_2O_3$ precipitates with the larger Young's modulus (400 GPa [17]) than $Y_2O_3$-stabilized $ZrO_2$ (210 GPa [18]) could increase the free energy of strain of monoclinic phase, and stability of the tetragonal phase consequently become high. This means that the critical stress to initiate the T→M phase transformation increases. According to the interpretation in Green *et al.*, the significant increase in $\sigma_f$ is explained by the increase in critical stress [16]. On the other hand, when the critical stress increases, the $K_{Ic}$ decreases because the T→M phase transformation would not be induced at the low stress. In addition, the $K_{Ic}$s of the specimens above were also evaluated by the single edge precracked beam (SEPB) method [19]. The $K_{Ic}$s were 10.4 MPa·m$^{0.5}$ for ZA-1350°C, 10.0 MPa·m$^{0.5}$ for ZA-1350°C-HIP, 10.3 MPa·m$^{0.5}$ for Z1A-1350°C-HIP, and 7.6 MPa·m$^{0.5}$ for Z20A-1500°C-HIP. Although the $K_{Ic}$ of each specimen was lower than that measured by the IF method due to the different measuring approaches, the tendency of difference between different samples was almost the same as that of the IF method.

Figure 3 presents an Ashby map showing $K_{Ic}$-$\sigma_f$ relationships for various engineering materials [20]. The $K_{Ic}$-$\sigma_f$ data obtained from ZA, Z1A, and Z20A were plotted in the map. These materials greatly exceeded the $K_{Ic}$-$\sigma_f$ range of technical ceramics including conventional zirconia and showed the $K_{Ic}$s comparable to those of metal-alloys and metallic glasses.

In summary, we successfully fabricated polycrystalline $ZrO_2$-based ceramics with high toughness, which greatly exceed the range of technical ceramics and are comparable to those of metallic materials, without any degradation of strength. The excellent mechanical properties in the proposed ceramic materials will further expand the application of advanced ceramics to many engineering fields.

**Methods**

**Specimen Preparation.** The starting powders were ZA (Zgaia 1.5Y-HT; Tosoh Co., Japan), Z1A (TZ-PX-714; Tosoh Co., Japan) produced by the refined hydrolysis process, and $Al_2O_3$ (TM-DAR; Taimei Chemicals Co., Ltd., Japan). The specific surface areas of ZA, Z1A, and $Al_2O_3$ powders were 15.6, 17.9, 14.4, and 14.5 $m^2/g$, respectively. The composite powder of Z20A was prepared by mixing the ZA and $Al_2O_3$ powders using a vibration mill. These powders were uniaxially pressed under a pressure of 70 MPa and were then cold-isostatically pressed at 200 MPa. The green compacts of ZA were sintered at temperatures in the range of 1000-1500°C at a heating rate of 100°C/h in air. The holding time at the set temperature was 2 h. The green compacts of ZA, Z1A, and Z20A were sintered at 1200, 1175, and 1350°C for 2 h at the same heating rate as above in air, and then were HIPed at 1350, 1350, and 1500°C for 1 h under a pressure of 150MPa in Ar, respectively. The $\rho$ of the sintered bodies was measured by the Archimedes method for $\rho$s greater than or equal to 80%, whereas it was calculated from the weights and outer dimensions of the sintered bodies for $\rho$s less than 80%.

**Characteristic evaluation.** The $\sigma_f$ was measured by the three-point-bending test according to JIS R1601. The measurement was performed by setting the distance between two supports (L) of 30 mm using a universal testing machine (Autograph AGS-X 10kN, Shimadzu, Japan) after prismatic specimens with 4 mm in width (w), 3 mm in thickness (t) and a length of at least 38 mm were prepared from the sintered body. The $\sigma_f$ was calculated using the following equation:

$$\sigma_f = \frac{3FL}{2wt^2} \tag{1}$$

where $F$ is the load at fracture. The $K_{Ic}$ of sintered bodies was determined by the IF method. Using a Vickers hardness tester (MV-1, Matsuzawa Co. Ltd., Japan), the indents on the surface of the specimen mirror-polished with 0.03μm-colloidal silica were created using a diamond indenter by applying a load of 196 N for 15 s. The radius ($c$) of the median crack and the length ($a$) of the half-diagonal of the indent were observed using a digital microscope (VHX-7000, Keyence, Japan). The $K_{Ic}$ was calculated using the equation reported by Lankford [21]:

$$\left(\frac{K_{Ic}\phi}{Ha^{1/2}}\right)\left(\frac{H}{E\phi}\right)^{0.4} = 0.142\left(\frac{c}{a}\right)^{-1.56} \tag{2}$$

where $H$ is the Vickers hardness, $E$ the Young's modulus, and $\phi$ the constraint factor. Here, the value of $\phi$ used 3 [25]. The value of $E$ used 210 GPa for ZA and Z1A and 250 GPa for Z20A [18].

**Microstructure Analysis.** The grain structure in the sintered bodies was observed by SEM (model JSM-7600F, JEOL, Japan). SEM specimens were mirror-polished with 1-μm diamond slurry and thermally etched for 2 min at a temperature 100°C lower than the sintering temperature of each specimen in air. The average grain size was measured by the planimetric method [22]. The specimen for TEM observation was mechanically ground to a thickness of approximately 50 μm, further dimpled to a depth of approximately 10 μm at its center, and then ion-milled to impart electron transparency. The microstructure and chemistry of the sintered bodies were investigated using STEM-EDS (JEM-ARM200CF, JEOL Co. Ltd) equipped with double silicon drift detector (SDD)-EDS detectors, of which the solid angle is about 1.7 sr.

**Figures**

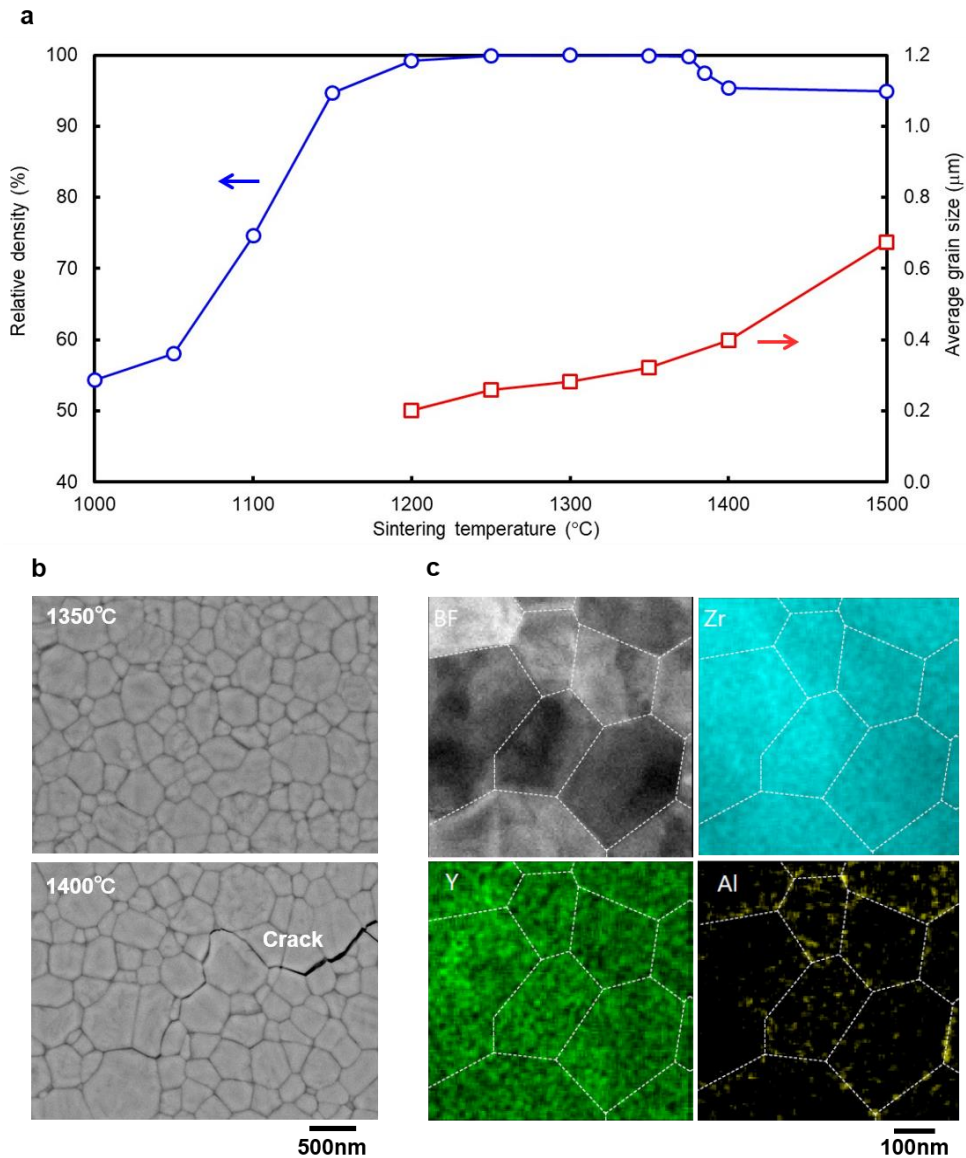

**Figure 1 | Densification grain-growth behavior and microstructure of fully densified bodies.**

**a**, Relative density and grain size of ZA as a function of their sintering temperature. **b**, SEM images of ZA sintered at 1350 and 1400°C. **c**, STEM image in ZA sintered at 1350°C and corresponding element maps for Zr-$K\alpha$, Y-$K\alpha$, and Al-$K\alpha$ acquired by the STEM-nanoprobe EDS technique.

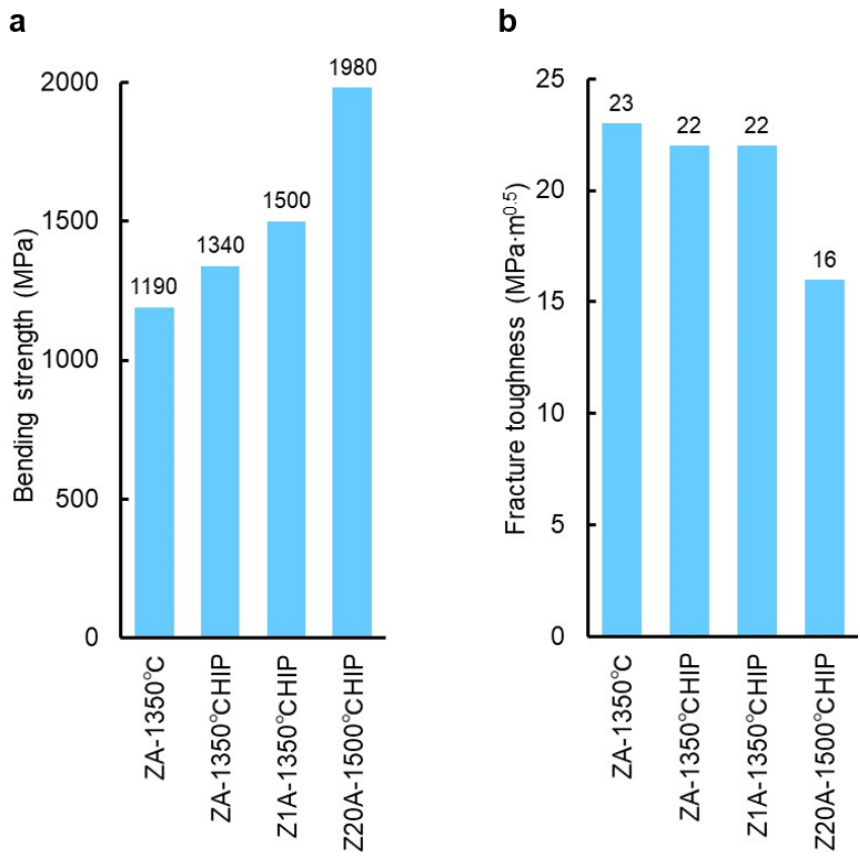

**Figure 2 | Mechanical properties. a**, Bending strengths and **b**, fracture toughnesses of ZAs sintered at 1250 and 1350°C, ZA and Z1A HIPed at 1350°C, and Z20A HIPed at 1500°C.

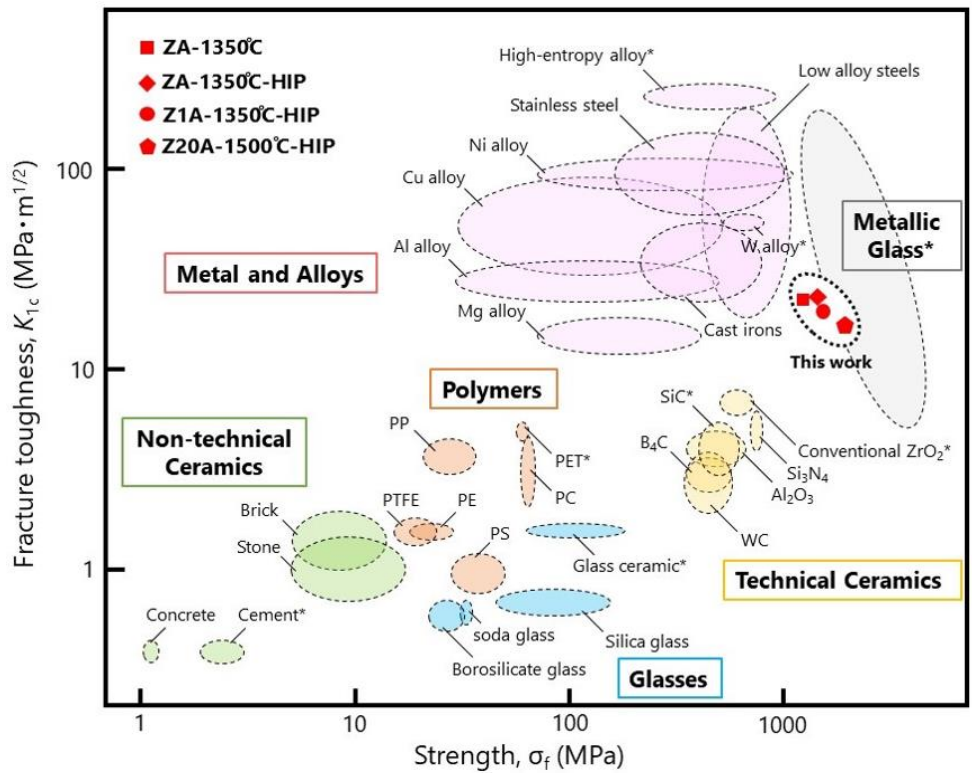

**Figure 3 | Ashby map showing fracture toughness-strength relationships for various engineering materials [18].** The materials marked with a star (*) are those cited from the report in Gludovatz *et al* [6]. The excellent fracture toughnesses and strengths of ZA, Z1A and Z20A greatly exceed those of technical ceramics and has a fracture toughness comparable to that of metal, alloy, and metallic glass.

**References**

1. Garvie, R. C., Hannink, R. H. & Pascoe, R. T. Ceramic steel?. Nature 258, 703-704 (1975).

2. Birkby, I. & Stevens, R. Applications of zirconia ceramics. Key. Eng. Mater. 122-124, 527-552 (1996).

3. Piconi, C. & Maccauro, G. Zirconia as a ceramic biomaterial. Biomater. 20, 1-25 (1999).

4. Kelly, J. R. & Denry, I. Stabilized zirconia as a structural ceramic: an overview. Dent. Mater. 24, 289-298 (2008).

5. Gupta, T. K., Lange, F. F. & Bechtold, J. H. Effect of stress-induced phase transformation on the properties of polycrystalline zirconia containing metastable tetragonal phase. J. Mater. Sci. 13, 1464-1470 (1978).

6. Gludovatz, B., Hohenwarter, A., Catoor, D., Chang, E. H., George, E. P. & Ritchie, R. O. A fracture-resistant high-entropy alloy for cryogenic applications. Science 345, 1153-1158 (2014).

7. Masaki, T. Mechanical properties of toughened $ZrO_2$-$Y_2O_3$ ceramics. J. Am. Ceram. Soc. 69, 638-640 (1986).

8. Matsui, K. Production method of zirconia powder for fine ceramics. J. Jpn. Soc. Powder Powder Metallurgy 68, 103-110 (2021).

9. Ii, S., Yoshida, H., Matsui, K., Ohmichi, N & Ikuhara, Y. Microstructure and surface segregation of 3 mol% $Y_2O_3$-doped $ZrO_2$ particles. J. Am. Ceram. Soc. 89, 2952-2955 (2006).

10. Tosoh Corporation, News Releases, December 6 (2021) https://www.tosoh.co.jp/news

/index.html

11. Tsukuma, T., Ueda, K. & Shimada, M. Strength and fracture toughness of isostatically hot-pressed composites of $Al_2O_3$ and $Y_2O_3$-partially-stabilized $ZrO_2$. J. Am. Ceram. Soc. 68, C4-C5 (1985).

12. Scott, H. G. Phase relationships in the zirconia-yttria system. J. Mater. Sci. 10, 1527-1535 (1975).

13. Matsui, K., Ohmichi, N., Ohgai, M., Yoshida, H. & Ikuhara, Y. Effect of alumina-doping on grain boundary segregation-induced phase transformation in yttria-stabilized tetragonal zirconia polycrystal. J. Mater. Res. 21, 2278-2289 (2006).

14. Matsui, K., Yamakawa, T., Uehara, M., Enomoto, N. & Hojo, J. Mechanism of alumina-enhanced sintering of fine zirconia powder: influence of alumina concentration on the initial stage sintering. J. Am. Ceram. Soc. 91, 1888-1897 (2008).

15. Tsukuma, K., Ueda, K., K. Matsushita, K. & Shimada, M. High-temperature strength and fracture toughness of $Y_2O_3$-partially-stabilized $ZrO_2/Al_2O_3$ composites. J. Am. Ceram. Soc. 68, C56-C58 (1985).

16. Green, D. J., Hannink, R. H. J. & Swain, M. V. in *Transformation toughening of ceramics* (eds Green, D. J.) 167-171 (CRC Press, Inc., Florida, 1989).

17. Asmani, M., Kermel, C., Leriche, A. & Ourak, M. Influence of porosity on Young's modulus and Poisson's ratio in alumina ceramics. J. Eur. Ceram. Soc. 21, 1081-1086 (2001).

18. Kondo, Y., Tsukuda, A. & Okada, S. Grindability of $Al_2O_3$-$ZrO_2$ composite ceramics. J.

Ceram. Soc. Jpn. 97, 929-934 (1989).

19. Nose, T. & Fuji, T. Evaluation of fracture toughness for ceramic materials by a single-edge-precracked-beam method. J. Am. Ceram. Soc. 71, 328-333 (1988).

20. Ashby, M. F. Materials Selection in Mechanical Design Ch. 3 (Pergamon, 1992).

21. Lankford, J. Indentation microstructure in the Palmqvist crack regime: implications for fracture toughness evaluation by the indentation method. J. Mater. Sci. 1, 493-495 (1982).

22. Yamaguchi, T. Characterization techniques of ceramic: properties of sintered bodies. Ceram. Jpn. 19, 520-529 (1984).

**Acknowledgments**

A part of this work was supported by the Nanotechnology Platform project, Grant Number JPMXP09A21UT0145, and the Elements Strategy Initiative for Structural Materials (ESISM), both of which are sponsored by MEXT, Japan.

**Author contribution statement**

K.M. and K.H. designed and conducted the experiments. B.F. and Y.I. conducted the STEM-EDS analyses. K.M. wrote the main manuscript text, and K.M. and B.F. prepared figures 1-4. All authors participated in discussions of the results and preparation of the manuscript.

**Additional information.**

Competing financial interests: The authors declare no competing financial interests.